\documentclass[12pt]{article}
\usepackage{cite}
\usepackage{array}

\usepackage{amssymb,bbold} %% remove if unavailable
\font\mybb=msbm10 at 11pt 
\def\bb#1{\hbox{\mybb#1}}

\def\bZ {\bb{Z}}
\def\bR {\bb{R}}

\def\bC {\bb{C}}
\def\CP {\bb{CP}}

\def\g{\gamma}

\def\ti{\tilde }

\newcommand{\R}{{\bR}}
\newcommand{\Z}{{\bZ}}
\newcommand{\C}{{\bC}}

\newcommand{\SO}{{\rm SO}}
\newcommand{\E}{{\rm E}}
\newcommand{\Sp}{{\rm Sp}}
\newcommand{\SL}{{\rm SL}}
\newcommand{\GL}{{\rm GL}}
\newcommand{\ISO}{{\rm ISO}}
\newcommand{\SU}{{\rm SU}}
\newcommand{\U}{{\rm U}}
\newcommand{\Spin}{{\rm Spin}}
\newcommand{\CSpin}{{\rm CSpin}}

\newcommand{\beq}{\begin{equation}}
\newcommand{\eeq}[1]{\label{#1}\end{equation}}
\newcommand{\bea}{\begin{eqnarray}}
\newcommand{\eea}[1]{\label{#1}\end{eqnarray}}

\newcommand{\fft}[2]{{\frac{#1}{#2}}}
\newcommand{\ft}[2]{{\textstyle{\frac{#1}{#2}}}}

  \ifx\ltimes\undefined
\newcommand{\ltimes}{{\kern3pt\hbox{\vrule width 0.4pt height 5.30pt  
depth
.0pt}\kern-1.76pt\times\kern1pt}}
\fi
\begin{document}
%%%%%%%%%%%%%%%%%%%%%%%%%%%%%%%%%%%%%%%%%%%%%%%%
\baselineskip 18pt

\begin{titlepage}

\vfill
\begin{flushright}
\today\\
QMUL-PH-03-03\\
hep-th/0305039\\
\end{flushright}

\vfill

\begin{center}
{\bf \Large
   Holonomy and  Symmetry in M-theory}

\vskip 10.mm {
Christopher M. Hull$^{*}$,}\\
\vskip 1cm

{\it
Department of Physics\\
Queen Mary, University of London\\
Mile End Rd, London E1 4NS, UK
}\\

\vspace{6pt}

\end{center}
\par

\begin{abstract}
\noindent  {
Supersymmetric solutions of 11-dimensional supergravity can be classified
according to the   holonomy of the supercovariant derivative arising in
the Killing spinor condition. It is shown that the holonomy must be
contained
 in $\SL(32,\R)$. The holonomies of  solutions with flux are discussed and 
 examples are analysed. In extending   to M-theory, account has to be
taken of 
 the phenomenon of \lq supersymmetry without supersymmetry'. It is argued
that 
 including the fermionic degrees of freedom in M-theory requires a
formulation with a
local $\SL(32,\R)$ symmetry, analogous to the need for local Lorentz
symmetry in coupling spinors to gravity.
}
\end{abstract}
\vskip 1cm
\vfill \vskip 5mm \hrule width 5.cm \vskip 5mm {\small
\noindent $^*$ E-mail: c.m.hull@qmul.ac.uk \\}
\end{titlepage}

%%%%%%%%

%%

%%%%%%%%%%%%%%%%%%%%%%%%%%%%%%%%%%%%%%%%%%%%%%%%
\section{Introduction}
\label{Intro}

In \cite{Duff:2003ec}, Duff and Liu addressed the two key questions of  
what the symmetries of M-theory might be, and how to  classify the  
supersymmetric solutions.
The concept of {\it generalised holonomy} played a central role in  
their discussion.
A bosonic solution of 11-dimensional supergravity will preserve $n$  
supersymmetries $(0\le n\le 32)$
if it admits $n$ spinor fields $\epsilon$ satisfying the condition
\begin{equation}
{\tilde D}_{M}\epsilon=0,
\end{equation}
  where ${\tilde D}_{M}$ is a certain connection (the supercovariant  
derivative) on the spin-bundle. The
number of solutions can then be analysed in terms of  the holonomy  
${\cal H} ({\tilde D})$ of the connection
  ${\tilde D}$. Such generalised  holonomy has been
  considered by a number of authors, including  
\cite{Duffstelle, Duff, Berkooz, Figueroa- 
O'Farrill:2002ft}.
   In \cite{Duff:2003ec}, the holonomy groups were analysed for a  
special class of warped product solutions   with a $d/(11-d)$ split in  
which the spacetime   decomposes locally into a
  $d$ dimensional piece and one of  $11-d$ dimensions.
  It was found that for splits in which the holonomy of the Levi-Civita  
connection $D$
  was in $\SO(d-1,1)\times \SO(11-d)$, the
  holonomy of $\tilde D$  was in the enlarged groups
   ${\cal G}=\SO(d-1,1)\times G_{spacelike}(11-d)$
   or ${\cal G}=G_{timelike}(d )\times \SO(11-d)$, where the groups
   $G_{spacelike}(n)$ and $G_{timelike}(n )$ are given in table 1.
   They also considered null splits in which the holonomy of the  
Levi-Civita connection
   is in $\ISO(d-1)\times \ISO (10-d)$, but that of $\tilde D$  is in
   ${\cal G}=\ISO(d-1)\times G_{null} (10-d)$, where the groups  
$G_{null} (n)$ are also given in
   table 1.

These same   groups ${\cal G}$ also arise as the local symmetry groups  
of 11-dimensional gravity dimensionally reduced on
a spacelike $n$-torus \cite{Cremmerjulia}, a timelike
  $n$-torus  \cite{Hulljulia} or a null $n$-torus  \cite{Duff:2003ec},  
respectively, for $n\le 8$.
For example, for a spacelike reduction on $T^n$, the
resulting theory has  local $\SO(d-1,1)$ Lorentz invariance where  
$d=11-n$, together with
a local $ G_{spacelike}(n)$ R-symmetry and a global $\E_{n(+n)}$  
duality symmetry
(where $\E_{n(+n)}$ is the maximally non-compact form of $\E_n$ and for  
$n\le 5$, the group is defined by a Dynkin diagram of the E-type, so  
that $\E_{5(+5)}=SO(5,5)$, $\E_{4(+4)}=\SL(5,\R)$ etc).
The scalar fields take values in $\E_{n(+n)}/ G_{spacelike}(n)$ and in  
the quantum theory the rigid symmetry is broken to a discrete subgroup
  $\E_{n(+n)}(\Z)$ \cite{Hull:1994ys}.
  For a timelike reduction, the theory has local $\SO(d )\times  
G_{timelike}(n )$ and global $\E_{n(+n)}$ symmetries, with scalars in  
$\E_{n(+n)}/ G_{timelike}(n)$ \cite{Hulljulia}.

  Remarkably, it has been shown that the full 11-dimensional  
supergravity can be rewritten in a form with local $\SO(d-1,1)\times  
G_{spacelike}(11-d)$
symmetry for $d=4$ \cite{Dewitnicolai2}, $d=3$ \cite{Nicolai}
  and $d=5,6$
  \cite{Drabant:1988bk}. These formulations involve   making a  
$d/(11-d)$ split and gauging away the  off-diagonal components of the  
vielbein.
  This led Duff and Liu to conjecture that
  there could be a similar formulation
  of the full 11-dimensional supergravity  theory using any of the  
groups ${\cal G} $, in which the
  field equations have a local ${\cal G} $ invariance.
  It could be the case that the spacetime symmetry group ${\cal G} $  
depends on certain features of the spacetime, such as whether it has a  
product structure, but it might also be the case that it is a larger  
group, containing all of the
  groups in table 1; the smallest such group is $\SL(32,\R)$. 
    This is a  
more interesting possibility, and could allow  a unified picture that  
is not background dependent.

  The purpose here is to consider the general  case in which no  
assumption is made about a product structure of the solution. The  
holonomy of $\tilde D$ must be
  contained in $\GL(32,\R)$ as 11-dimensional  Majorana spinors    are  
real and have 32 components.
  Moreover, it should contain the groups in table 1, so in particular   
it should contain both $\SO(32)$ and $\SO(16,16)$. As will be seen,   
the holonomy of $\tilde D$  must in fact be in
   $\SL(32,\R)$, and the consequences of this for supersymmetric  
solutions will be explored. The holonomy is   $\SL(32,\R)$for generic
backgrounds, and particular classes of background
    have holonomy restricted to special subgroups, such as backgrounds
with a $d/(11-d)$ split which   have the special holonomies in table 1.
Examples will be considered in which the holonomy is in  
other subgroups of $\SL(32,\R)$ that did not arise in   
\cite{Duff:2003ec}, e.g for static solutions with electric flux, the  
holonomy is in $\SL(16,\C)$ or, with an additional assumption on the  
ansatz, in $\Spin(10,\C)$.

   This, together with the arguments of
    \cite{Duff:2003ec},
    motivates the conjecture that there should be a formulation of  
11-dimensional
  supergravity in which there is a local  $\SL(32,\R)$ symmetry. This  
would be similar to
  the formulations of  \cite{Dewitnicolai2},  \cite{Nicolai}
  and
  \cite{Drabant:1988bk} and in this sense it could then be said that
  the 11-dimensional theory would have a hidden $\SL(32,\R)$ spacetime  
symmetry.

  %%%%%%%%%%%%%%%%%%%%%%%%%%%%%%%%%%%%%%%%%%%%%%%%%%%%%%%%%%%%%%%%%%
\begin{table}[t]
\begin{center}
\begin{tabular}{c|lll}
$n$&$ G_{spacelike}(n)$&$G_{timelike}(n) $&$G_{null}(n) $\\
\hline
1&$ \{1\}$ & $ \{1\}$ &$ \{1\}$\\
2&$  \SO(2)$&$ \SO(1,1)$&$ \R$\\
3&$ \SO(3) \times \SO(2)$&$ \SO(2,1) \times  \SO(1,1)$&$ \ISO(2) \times  
\R$\\
4&$ \SO(5)$&$\SO(3,2)$&$ [\SO(3) \times \SO(2)] \ltimes
\R^6_{(3,2)}$\\
5&$ \SO(5) \times \SO(5)$&$ \SO(5,\C)$&$ \SO(5) \ltimes  
\R^{10}_{(10)}$\\
6&$ {\rm USp}(8)$&${\rm USp}(4,4)$&$ [\SO(5) \times \SO(5)] \ltimes
\R^{16}_{(4,4)}$\\
7&$ \SU(8)$&$\SU^*(8)$&${\rm USp}(8)\ltimes \R^{27}_{(27)}$\\
8&$ \SO(16)$&$ \SO^*(16)$&$ [\SU(8) \times \U(1)]\ltimes
                         \R^{56}_{(28_{1/2},\overline{28}_{-1/2})}$\\
9&$ \SO(16) \times \SO(16)$&$ \SO(16,\C)$&$ \SO(16) \ltimes  
\R^{120}_{(120)}$\\
10&$\SO(32)$&$\SO(16,16)$&$  
[\SO(16)\times\SO(16)]\ltimes\R^{256}_{(16,16)}$
\end{tabular}
\end{center}
\caption{Generalized structure groups. For spacelike reductions, the  
holonomy group is in
${\cal G}=\SO(d-1,1)\times G_{spacelike}(11-d)$,
   for timelike ones it is in  ${\cal G}=G_{timelike}(d-1,1)\times  
\SO(11-d)$ while for null ones it is in
   ${\cal G}=\ISO(d-1)\times G_{null} (10-d)$.}
\label{gen}
\end{table}
%%%%%%%%%%%%%%%%%%%%%%%%%%%%%%%%%%%%%%%%%%%%%%%%%%%%%%%%%%%%%%%%%%

An important issue regarding such reformulations of $D=11$ supergravity  
is that
there is a sense in which they are simply rewritings of the original
theory and have no  
physical content. After all, it is   possible to enlarge the  
symmetry of any theory by introducing extra degrees of freedom, and then
introducing extra symmetries that can be used to eliminate these  extra  
degrees of freedom.
For the dimensionally reduced supergravity theory in $11-n$ dimensions  
($n\le 8$), the physical scalars take values in the coset  
$E_{n(+n)}/G(n)$ and the theory has a non-linearly realised
$E_{n(+n)}$ global symmetry, where $G(n)$ is the appropriate group from  
table 1.
Introducing extra scalars taking values in the group $G(n)$
leads to a formulation in which the $E_{n(+n)}$  global symmetry is  
linearly realised,
and in which there is a new local $G(n)$ symmetry which can be used to  
gauge away the
extra unphysical scalars  \cite{Cremmerjulia}.
This local $G(n)$ symmetry  is not an essential part of the classical  
theory, but it is very convenient to write the theory in a formulation  
with this symmetry.

As was to be expected, the situation is similar in the $ 11  
$-dimensional  formulations with local
$G(n)$ symmetry.
In
   \cite{Dewitnicolai2},   \cite{Nicolai}
  and   \cite{Drabant:1988bk}, extra fields
  % taking values in the group $G(n)$
   are introduced in $d=11$ which can be gauged away by the local
  $G(n)$ symmetry in $d=11$. In both the reduced and eleven-dimensional  
supergravities,
  introducing the extra fields and extra $G(n)$ symmetry is a matter of  
convenience
  leading to a useful way of formulating the theory, but the local  
symmetry is not
an essential part of the theory, although it is suggestive that many   
of the interactions have such a symmetry.

An important question for M-theory then is whether the symmetry $G(n)$  
is a convenience leading to a useful way of formulating the theory, as  
in supergravity, or whether it is an essential part of the theory.
It will be argued here that for M-theory    extra symmetries such as those
in    
table 1 play a crucial role and that they do act non-trivially on  
physical degrees of freedom, so that the theory cannot be written in a  
form without these symmetries.
In particular, it will be seen that certain physical degrees of freedom  
of M-theory arise as sections of bundles with
transition functions in the structure group ${\cal G} $ and which  
cannot be regarded as
sections of e.g. the spin bundle. These arise in situations where there  
is \lq supersymmetry without supersymmetry'  
\cite{DLP1,DLP2,Pope:1999xg}, corresponding to M-theory vacua which are  
known to be supersymmetric but for which
the corresponding supergravity solution does not have Killing spinors.

It is interesting to compare with gravity. General relativity in
$d$-dimensions can be formulated in terms of a metric, and the holonomy is
in $SO(d-1,1)$. It can instead be formulated in terms of a vielbein with
local $SO(d-1,1)$ Lorentz symmetry. This involves introducing extra fields
(the extra components of the vielbein)    together with an  extra  local
Lorentz gauge  symmetry that can be used to eliminate them, so that the
number of degrees of freedom remains the same.
For pure gravity, this is just a convenient rewriting of the theory with a
tangent space group that is the same as the holonomy group.
However, for coupling to spinor fields, the formulation with local Lorentz
symmetry is essential. 
A similar story seems to be true for M-theory.
In classical supergravity (reduced from 11 to $d$ dimensions), one can
work in physical gauge (with no extra scalars) and the generalised
holonomy is in the appropriate group ${\cal G}$ from table 1, 
or one can introduce extra scalars so that the structure group ${\cal G}$
is the same as the holonomy group; the two formulations are equivalent.
However, it will be argued that 
to describe the fermionic degrees of freedom in M-theory, local ${\cal G}$
symmetry is essential and so a formulation with such symmetry is required.
A formulation with local $\SL(32,\R)$ symmetry allows the coupling to all
such degrees of freedom that can arise, and  the extra symmetry is
independent of the choice of background.

The plan of the paper is as follows. In section 2, the generalised  
holonomy of $\ti D$ will be reviewed, and it will be shown that in  
general it is in $\SL(32,\R)$. In section 3, the holonomy will be  
discussed further and the number of supersymmetries will be discussed.  
In section 4, the holonomies of certain examples will be considered.  
Section 5 extends the discussion to
  other 11-dimensional supergravities.
  Section 6 discusses M-theory, and in particular the phenomenon of  
supersymmetry without supersymmetry, and argues that M-theory requires  
fermions which are sections of an $\SL(32,\R)$ bundle, rather than the  
spin bundle.
It concludes with some speculative remarks. Further details 
of the structure groups that arise in gauged supergravities
are given in an appendix.
  %%%%%%%%%%%%%
%%%%%%%%%%%%%%%%%%%%%%%%%%%%%%%%%%%%%%%%%%%%%%%%%%

\section{Killing Spinors and Generalized Holonomy}
\label{holonomy}

The  fields of $D=11$ supergravity    are a graviton $g_{MN}$, a  
gravitino $\Psi_M$ and $3$-form gauge field
$A_{MNP}$, where $M=0, 1, \ldots 10$. All spinors are in the Majorana  
spinor representation of $\Spin(10,1)$, and a vielbein $e_M{}^A$ is  
used to convert coordinate indices
$M,N$ to tangent space indices $A,B$.
The bosonic field equations   (setting $\Psi_M=0$) are
\begin{equation}
R_{MN}=\frac{1}{12}\left(F_{MPQR}F_{N}{}^{PQR}-\frac{1}{12}g_{MN}
F^{PQRS}F_{PQRS}\right)
\end{equation}
and
\begin{equation}
d*\!F_{(4)}+\fft12F_{(4)} \wedge F_{(4)}=0,
\end{equation}
where $F_{(4)}=dA_{(3)}$.  The supersymmetry transformation rule of the
gravitino  in a   bosonic background is
\begin{equation}
\delta \Psi_{M}={\tilde D}_{M} \epsilon,
\label{eq:11gto}
\end{equation}
with spinor    parameter $\epsilon$, where
\begin{equation}
{\tilde D}_{M}=D_{M}-
\frac{1}{288}(\Gamma_M{}^{NPQR}-8\delta_M^N\Gamma^{PQR})F_{NPQR},
\label{supercovariant}
\end{equation}
The  $\Gamma_{A}$ are   $D=11$ Dirac matrices and $\Gamma_{AB...C}$ are  
antisymmetrised products of gamma matrices,  
$\Gamma_{AB...C}=\Gamma_{[A}\Gamma_{B}...\Gamma_{C]}
$. The signature is $(-++\dots+)$ and a Majorana representation is used  
in which
the  spinors have 32 real components and the gamma-matrices are real.
Here $D_{M}$ is the
usual Riemannian covariant derivative involving the Levi-Civita
connection  
$\omega_{M}$
taking values in the tangent space  group $\Spin(10,1)$
\begin{equation}
D_{M}=\partial_{M}+\frac{1}{4}\omega_{M}{}^{AB}\Gamma_{AB}.
\end{equation}
In the quantum theory, the field equations and supersymmetry  
transformations receive higher derivative corrections; these will not  
be considered explicitly here.
Note that a space admitting Killing spinors does not necessarily  
satisfy the field equations.

Each solution of
\begin{equation}
{\tilde D}_{M}\epsilon=0,
  \label{supercovariantaa}
\end{equation}
is a Killing spinor field that
generates a supersymmetry leaving the background invariant, so that
the number of supersymmetries preserved by a supergravity background  
depends
on the number of supercovariantly constant spinors satisfying   
(\ref{supercovariantaa}).
Any commuting Killing spinor field $\epsilon$ defines a Killing vector  
$v_{A}=\overline{\epsilon}\Gamma_{A}\epsilon$, which is either timelike  
or null, together with a 2-form $\overline{\epsilon}\Gamma_{AB}\epsilon$
and a 5-form $\overline{\epsilon}\Gamma_{ABCDE}\epsilon$.

If $F_{(4)}=0$, then $\ti D=D$ and the Killing spinors are covariantly  
constant with respect to the
Levi-Civita connection.
If the holonomy group of this connection is ${\cal H}(D)\subseteq  
\Spin(10,1)$, then the covariantly constant spinors
   are the  singlets of the holonomy group ${\cal H}(D)$
under the decomposition of the   ${\bf 32}$ of $\Spin(10,1)$ under  
${\cal H}(D)$.
For Euclidean signature, the
holonomy groups have been classified by Berger \cite{Berger}, while in
Lorentzian  
signature the
holonomies of spacetimes with parallel spinors were analysed by Bryant  
\cite{Bryant}.
Suppose there is at least one Killing spinor $\epsilon$. If
the   Killing vector $v_{A}=\overline{\epsilon}\Gamma_{A}\epsilon$
is timelike, then ${\cal H}(D)\subseteq SU(5) \subset
\Spin(10,1)$ and the allowed values for the number of Killing spinors
are  2, 4, 6, 8,
16, 32 \cite{Acharya:1998yv,Acharya:1998st}.  If on the other hand
the   Killing vector $v_A$
is null, then ${\cal H}(D)\subseteq  \R\times(\Spin(7) \ltimes \R^8)   
\subset
\Spin(10,1)$ and the allowed values for the number of Killing spinors
are   1, 2, 3, 4, 8, 16, 32   \cite{Acharya:1998yv,Acharya:1998st}.

If $F_{(4)}\neq 0$, then $\ti D$ is a connection on the spin bundle.
The  Clifford algebra $Cl(10,1)$ is spanned by the matrices
$\{ 1,  
\Gamma_{A},\Gamma_{AB},\Gamma_{ABC},\Gamma_{ABCD},\Gamma_{ABCDE}\}$
and is the algebra of real $32\times 32 $ matrices, $Mat(32,\R)$. In  
particular, the commutation relations of these matrices are those of   
the algebra $GL(32,\R)$, and   the holonomy  of $\ti D$ must   be  
contained in $GL(32,\R)$.
Note that $\Gamma_{AB}$ generate $\Spin(10,1)$,
$ \{ \Gamma_{A},\Gamma_{AB} \}$ generate the subalgebra
$\Spin(10,2)$ and
$ \{  \Gamma_{A},\Gamma_{AB} ,\Gamma_{ABCDE} \}$ generate the subalgebra
$\Sp(32,\R)$.

In  \cite{Duff:2003ec}, spaces
with a product structure were considered,
with Riemannian holonomy
${\cal H}(D)\subseteq \SO(d-1,1)\times \SO(11-d)$,
  allowing a
  $d/(11-d)$
split. Attention was restricted to cases that allow a dimensional
reduction to $d$ dimensions. The system was truncated to one in which only
the metric  
and scalars in $d$ dimensions were kept, with the ansatz
\begin{equation}
g^{(11)}_{MN}=\pmatrix{\Delta^{-1/(d-2)}g_{\mu\nu}&0\cr0&g_{ij}},\qquad
A^{(11)}_{ijk}=\phi_{ijk},
\label{ansatz}
\end{equation}
where $\Delta=\det{g_{ij}}$.
The $d$-dimensional  fermion fields were defined as
\begin{equation}
\psi_\mu=\Delta^{\fft1{4(d-2)}}\left(\Psi^{(11)}_\mu+\fft1{d- 
2}\gamma_\mu
\Gamma^i\Psi^{(11)}_i\right),\qquad
\lambda_i=\Delta^{\fft1{4(d-2)}}\Psi^{(11)}_i,
\label{siis}
\end{equation}
with the gravitino transforming as
\begin{equation}
\delta\psi_\mu=\hat D_\mu\epsilon,\qquad
\hat D_\mu=\partial_\mu+\ft14\Omega_\mu,
\end{equation}
with a certain generalised connection $\Omega_\mu$.
The supersymmetry was analysed in  \cite{Duff:2003ec} in terms of the  
holonomy of the supercovariant derivative $\hat D_\mu $
of the reduced system, and an important role was played by
the fact that
   $\Omega_\mu$ involves only
$\gamma_{\alpha\beta}$ together with the algebra generated by   
$\{\tilde \Gamma_{ab}, \tilde\Gamma_{abc} \}$.
Here $\gamma_\alpha$ are $\SO(d-1,1)$ Dirac matrices, while  
$\tilde\Gamma_a$
are $\SO(11-d)$ Dirac matrices.
The holonomy group is then $\SO(d-1,1)\times G_{spacelike}(11-d)$ where  
$G_{spacelike}(11-d)$ is the group generated by
$\{ \tilde\Gamma^{(2)},\tilde\Gamma^{(3)},  
\tilde\Gamma^{(6)},\tilde\Gamma^{(7)}, \tilde\Gamma^{(10)}\}$,
with the notation that $\tilde\Gamma^{(n)}$ represents the  
antisymmetric product of
$n$ gamma matrices, $\tilde\Gamma_{a_1....a_n}$. (For $n>11-d$,  
$\tilde\Gamma^{(n)}=0$.)
This gives the groups in table 1, with the exception that for $d=4$,  
the generator $\tilde\Gamma^{(7)}$ which could have occured in the  
algebra in fact does not arise on the right hand side of any  
commutators of the algebra generated by  
$\{\tilde\Gamma^{(2)},\tilde\Gamma^{(3)}, \tilde\Gamma^{(6)}\}$, so  
that the algebra is $G_{spacelike}(7)=SU(8)$ rather than the $U(8)$  
that would have arisen on adding $\tilde\Gamma^{(7)}$.

The definition of $\psi_\mu$ in (\ref{siis}) eliminates the terms  
involving $ \Gamma^{(5)}$ in $\delta\Psi^{(11)}$ from $\delta  
\psi_\mu$, which now appear
in $\delta \lambda ^i$. The conditions for supersymmetry considered in   
\cite{Duff:2003ec} that there be spinors
satisfying $\hat D_\mu \epsilon=0 $, 
  restricting the holonomy of $\hat D_\mu$, are then necessary  
but not sufficient, as they
must be supplemented by the conditions $\delta \lambda ^i=0$, and the  
holonomy of
$\hat D $ is in general different from that of
$\tilde D $. Here, the emphasis will  be   on the
holonomy of $\tilde D $, giving necessary and sufficient conditions for  
supersymmetry.
Examples will be considered in the next section.

  In general, from the form of $\ti D$, the holonomy for $\ti D$ must be  
in the subalgebra of
  $\GL(32,\R)$ generated by
  $\{ \Gamma^{(2)},\Gamma^{(3)}, \Gamma^{(5)}\}$.
  From above, closing the algebra  generated by $\{
\Gamma^{(2)},\Gamma^{(3)}\}$
leads to the set $\{\Gamma^{(2)},\Gamma^{(3)}, \Gamma^{(6)},\Gamma^{(7)},  
\Gamma^{(10)}\}$ which,
using the fact that
$\Gamma^{(n)} \propto *  
\Gamma^{(11-n)} $,
  is the algebra
  generated by
  $\{ \Gamma^{(1)},\Gamma^{(2)},\Gamma^{(3)},  
\Gamma^{(4)},\Gamma^{(5)}\} $. In particular, adding $\Gamma^{(5)}$
to this does not enlarge the algebra.
The issue is then whether $\Gamma^{(11)} = 1$   occurs on the right  
hand side of any commutators. A calculation shows  that it does not  
(the situation is similar to the absence of $\Gamma^{(7)}$
for the case $11-d=7$
discussed above) so that the algebra is
indeed $\SL(32,\R)$, not the full $\GL(32,\R)$.
  As a result, the holonomy of  $\ti D$ must be contained in  
$\SL(32,\R)$.

%%%%%%%%%%%%%%%%%%
%%%%%%%%%%%%%%%%%%%%%%%%%%%%%%%%

\section{Supersymmetric Backgrounds and Special Holonomies}
\label{susy}

\subsection{Holonomies and Structures}

  The key question is   which subgroups of $\SL(32,\R)$ actually occur  
as holonomies of supergravity backgrounds. This is the analogue of the  
question of which holonomies can occur for Riemannian manifolds, which  
was answered by Berger  \cite{Berger}.

  It is often useful to write the supercovariant derivative
  as
  \begin{equation}
\label{fghg}
\ti D_M = \hat D_M + X_M
\end{equation}
  for some other connection  $\hat D_M$ on the spin-bundle, and some  
covariant $32\times 32 $ matrix $X_M$.
  Then one can make the ansatz in which one  seeks backgrounds admitting
Killing spinors satisfying the  
algebraic constraints
\begin{equation}
\label{qwer}
X_M \epsilon=0
\end{equation}
These should also satisfy $\hat D_M\epsilon =0$, and so can be analysed  
in terms of the holonomy of the {\it associated dertivative} $\hat D_M$.
Clearly ${\cal H}(\hat  
D_M)\subseteq {\cal H}(\ti D_M)$, but often
   ${\cal H}(\hat D_M)$ is easier to analyse.
   For example, as reviewed in the last section, Duff and Liu  analysed  
the holonomy of
   the connection $\hat D$ arising from requiring $\delta \psi _\mu=0$  
where $\psi_\mu$ is defined in (\ref{siis}), which must be supplemented by
the  
condition $\delta\lambda_i=0$, which is algebraic of the form (\ref{qwer})
for their ansatz.

   As another example, consider the case in which $X_M = \Gamma_M f$  
where
\begin{equation}
\label{fis}
f= \frac{1}{24}F_{MPQR} \Gamma^{MPQR}
\end{equation}
and note that the derivative (\ref{supercovariant}) can be rewritten as
\begin{equation}
{\tilde D}_{M}=D_{M}+ \frac{1}{24} \Gamma^{PQR}F_{MPQR}
- \frac{1}{12}\Gamma_Mf
  \label{supercovarianta}
\end{equation}
Then for backgrounds in which   the Killing spinor satisfies
\begin{equation}
\label{fiso}
f\epsilon  = 0\end{equation}
the Killing spinor condition simplifies to
\begin{equation}
{\hat D}_{M}\epsilon \equiv (D_{M}+ \frac{1}{24}  
\Gamma^{PQR}F_{MPQR})\epsilon=0
  \label{kil}
\end{equation}
and the   analysis of supersymmetric backgrounds in terms of the holonomy  
${\cal H}(\hat D)$
will be explored in the next section.

It will be useful to refer to the maximal holonomy group for a class of
configurations as the structure group.
Thus the structure group associated with $\ti D$ for a generic
configuration is the group $\SL(32,\R)$ generated by
$\{ \Gamma^{(2)},\Gamma^{(3)}, \Gamma^{(5)}\}$.
As we have seen, for special classes of configuration, the structure group
for  a related operator $\hat D$
is the group generated by $\{ \Gamma^{(2)},\Gamma^{(3)}\}$. In 11
dimensions, 
this is the same group  $\SL(32,\R)$, but in lower dimensions or for
product spaces, this leads
to the groups in table 1, as will be explored further in the next section.
The particular subgroup of the structure group that arises as the holonomy
will determine the number of Killing spinors.

Note that $\ti D$  is not the most general $\SL(32,\R)$ 
connection one could write down, because of the particular way that $F$
enters into the expression, so in
principle it could have been
that it would have led to a structure group smaller than 
$\SL(32,\R)$. However, as the structure group has to contain both
$\SO(32)$ and  $\SO(16,16)$, 
together with the other groups found in the next section, the
structure group must in fact be $\SL(32,\R)$.

  The connection $\ti D$  on the spin bundle extends to tensor products,  
so that one can define the
  supercovariant derivative of multi-spinors  
$\chi^{\alpha\beta....\gamma}$.
  The only invariant of $\SL(32,\R)$ is the 32nd rank alternating  
tensor, while for the subgroup
  $\Sp(32,\R)$ there is an invariant anti-symmetric 2-form  
$C^{\alpha\beta}$, the charge conjugation matrix. For the subgroup   
$\Spin(10,1)$ (or $\Spin(10,2)$, $\Spin(6,5)$, $\Spin(6,6)$) there
  are invariant gamma matrices $(\Gamma_M)^\alpha{}_\beta$.
A bi-spinor can be related to a set of forms using gamma-matrices, so  
that
   $\chi^{\alpha\beta }$ can be written as a linear combination of the
   $n$-forms $\chi  
_{M_1...M_n}=(C^{-1}\Gamma_{M_1...M_n})_{\alpha\beta}\chi^{\alpha\beta }$  
for $n=0,1,2,3,4,5$, where $C$ is the charge conjugation matrix. If  
${\cal H}(\ti D)$ is in $\Spin(10,1)$, then there is a natural  
extension of  $\ti D$ to a metric connection   on the tangent bundle so  
that the
   gamma-matrices are  supercovariantly constant, and a bi-spinor  
satisfying
    $\ti D\chi^{\alpha\beta }=0$
  will then define forms that are  supercovariantly constant, $\ti D\chi  
_{M_1...M_n}=0$.
  In this case, one can consider $\ti D$ as a connection on the tangent  
bundle, and analyse its
  holonomy.
  However, for holonomies not in
  $\Spin(10,1)$ (or one of the other $\Spin$ subgroups)  there is no  
natural definition of  $\ti D$
    on the tangent bundle.
    Nonetheless,  if a space admits Killing spinors, then the tangent
bundle  
will have a G-structure, i.e. it can be regarded as a bundle with  
transition functions in some group $G\subset \Spin(10,1)$, with
    the group related to the number of Killing spinors, whcih are
singletsd under $G$; this has been  
used to analyse the geometry associated with Killing spinors in  
\cite{Gauntlett:2002fz}.
There are many more subgroups of $\SL(32,\R)$ that can arise as holonomies
than there are subgroups
$G\subset \Spin(10,1)$ that can arise in G-structures; for example, 
all of the solutions with $16 <n\le 32 $ supersymmetries have trivial
G-structures
(with G=1) but each of the different values of $n$ corresponds to a
different holonomy.
Then the generalised holonomy may be more useful in classifying 
supersymmetric spaces, while the G-structure approach of
\cite{Gauntlett:2002fz} 
is more useful in the construction of explicit  solutions.

\subsection{Number of Supersymmetries}

The superalgebra in 11-dimensions with tensorial charges  
\cite{Hull:1997kt}
allows any number  of supersymmetries $0\le n\le 32$
to be preserved by a state \cite{Gauntletthull1}.
This is a non-trivial statement, as other superalgebras in other  
dimensions place restrictions on the allowed number of supersymmetries.
Many of the supersymmetric solutions are not asymptotically flat, and  
so are difficult to  analyse in terms of a global superalgebra.
For this reason, it seems more useful to address the
problem through the Killing spinor equation and generalised holonomy.

Until recently,  no solutions with $16< n < 32$  supersymmetries were  
known, but now   solutions   are   known preserving
0, 1, 2, 3, 4, 5, 6, 8, 10, 12, 14, 16, 18, 20, 22, 24,
26, 28, 32 supersymmetries \cite{Ohta} --\cite{Harmark}.
These values can   be discussed in terms of the holonomy groups of  
table 1 \cite{Duff:2003ec}.
(Note that the conditions for supersymmetry from the holonomy of $\hat  
D$ are necessary but not sufficient in \cite{Duff:2003ec}, as one also  
needs to take into account the conditions from the supersymmetry  
variations of the spin-half fermions arising in the dimensional  
reduction.)

In \cite{Bandos:2001pu}, it was suggested that there could be \lq  
preon' states in M-theory preserving 31 supersymmetries
from which all other BPS states might arise as bound states.
Although their conjecture does not require that there be classical 
supergravity solutions preserving 31 supersymmetries, it is nonetheless
interesting to ask whether such backgrounds exist.
In \cite{Duff:2003ec}, it was shown that $n=31$ cannot arise from  
holonomies contained in any of the structure groups in table 1.
However, it could be that such solutions could occur
without being of the type arising from    the ansatz of   
\cite{Duff:2003ec}.

For a configuration preserving $n$ supersymmetries, the holonomy must  
be in
$\SL(32-n,\R)\ltimes \R^{n(32-n)}$. All values of $n$ are then possible in
principle, including $n=31$, which would require a non-trivial  
$\R^{31}$ holonomy.
The issue is then whether this holonomy can actually arise from a
supergravity configuration.
The maximally supersymmetric solutions have been classified in  
\cite{Figueroa-O'Farrill:2002ft}.

%%%%%%%%%%%%%%%
\section{Examples of Supersymmetric Backgrounds and their Holonomies}
\subsection{Examples with Cosmological Constant}

Consider the Freund-Rubin ansatz of a product space  with a 4/7 split  
in which the holonomy group of the Levi-Civita connection is in  
$\Spin(3,1)\times
\Spin(7)$ and with
  \begin{equation}
F_{\mu\nu\rho\sigma}=3\mu \epsilon_{\mu\nu\rho\sigma}
\end{equation}
where $\mu,\nu,\rho=0,1,2,3$ and $\mu$ is a constant with the dimensions
of mass.
Then the space is  the product of a four-dimensional spacetime with  
negative
curvature
\begin{equation}
R_{\mu\nu}=-12\mu^{2}g_{\mu\nu}
\end{equation}
and a seven-dimensional   space of positive curvature
\begin{equation}
R_{ij}=6\mu^{2}g_{ij}
\end{equation}
where $i,j=1,2,\ldots 7$.
The supercovariant derivative is then
given by
\begin{equation}
{\tilde D}_{\mu}= D_{\mu}+im\gamma_{\mu}\gamma_{5}
\end{equation}
and
\begin{equation}
{\tilde D}_{i}= D_{i}-\frac{i}{2}\mu\Gamma_{i}
\end{equation}
The holonomy of these two connections are $\Spin (3,2)$ and $\Spin(8)$  
respectively, so in this case the structure group is $\Spin (3,2)  
\times \Spin(8)$
and the holonomy must be a subgroup of this. Note that both factors in  
$\Spin(3,1)\times
\Spin(7)$ have been enlarged. The maximally supersymmetric solution  
$AdS_4\times S^7$ has trivial holonomy.
A similar ansatz with a product of a Euclidean four-dimensional space  
and a Lorentzian seven-dimensional one has the Riemannian structure  
group
$\Spin(4)\times \Spin(6,1)$ enhanced to
$\Spin(5)\times \Spin(6,2)$, with maximally supersymmetric solution  
$AdS_7\times S^4$ with trivial holonomy.

This is an example of something that occurs in other contexts and so
it is worth considering  more generally.
Consider first a $d$ dimensional space  $X$ with
positive definite metric, so that ${\cal H} (\omega) \subseteq  
\Spin(d)$,  and consider Killing spinors
satisfying
\begin{equation}
{\tilde D}_{i}\epsilon = 0
\end{equation}
where the derivative is
\begin{equation}
{\tilde D}_{i}= D_{i}-im\Gamma_{i} \quad ~ {\rm if} ~ d~ {\rm odd},
\qquad
{\tilde D}_{i}= D_{i}- m\Gamma_{i}\Gamma_{*}
\quad ~ {\rm if} ~ d~ {\rm even}
\label{conkil}
\end{equation}
  where $\Gamma_{*}$ is the chirality operator in $d$ dimensions,  
$\Gamma_{*}\propto \prod _i \Gamma_{i}$.
  The geometries that give rise to different number of spinors have been  
classified in \cite{Acharya:1998db}.
As $i\Gamma_{i}, \Gamma_{ij}$  together generate $\Spin(d+1)$ if $d$ is  
odd,
while $\Gamma_{i}\Gamma_{*}, \Gamma_{ij}$ generate $\Spin(d+1)$ if $d$  
is even,
  the holonomy
of ${\tilde D}_{i}$ will in either case be in  the structure group
$\Spin(d+1)$, and the
  number of
Killing spinors will depend on this
holonomy. The geometries arising for various holonomies ${\cal H}(\ti  
D)\subset \Spin(d+1)$ are given in table 2, and these holonomies are a
useful way of categorising these spaces.  For odd $d$, the number of  
Killing spinors depends on the sign of $m$, and the   table gives the  
numbers $n_+,n_-$ of Killing spinors
for either sign. For odd $d$, changing the orientation of $X$  
effectively changes the sign of
$m$.

\begin{table}[h!]
\centering
\setlength{\extrarowheight}{3pt}
\begin{tabular}{|>{$}c<{$}|>{$}c<{$}|c|>{$}c<{$}|}\hline
d=\dim X &  {\rm Generalised ~Holonomy } & Geometry of $X$ & (n_+,  
n_-)\\
\hline
\hline
d & \{1\} & round sphere & (2^{\lfloor d/2\rfloor},2^{\lfloor
d/2\rfloor})\\
4k-1 &\Sp(k)& $3$-Sasaki & (k+1,0)\\
4k-1 &\SU({2k}) & Sasaki--Einstein & (2,0)\\
4k+1 &\SU({2k+1})& Sasaki--Einstein & (1,1)\\
6 & G_2 & nearly K\"ahler & (1,1)\\
7 & \Spin (7) & weak $G_2$ holonomy & (1,0)\\[3pt]
\hline
\end{tabular}
\vspace{8pt}
\caption{Table 2.Manifolds admitting real Killing spinors}
\label{tab:geometries}
\end{table}

The holonomy ${\cal H}_X(\ti D)$ is also the holonomy of the  
Levi-Civita connection ${\cal H}_{C(X)}(  D)$
on the cone over $X$, $C(X)$,
which has metric $ds^2_{C(X)}=dr^2 + r^2 ds^2_X$ where $ds^2_X$ is the  
metric on $X$.
For example the cone over a 3-Sasaki space
is hyperk\" ahler.
The round sphere $S^d$ with isometry $SO(d+1)$ has maximal  
supersymmetry and trivial   holonomy  ${\cal H}(\ti D)$, and the cone  
$C(S^d)=\R^{d+1}$ is
flat space with ${\cal H}( D)=1$.

If $ m$ is replaced with $im$ in (\ref{conkil}), so that
\begin{equation}
{\tilde D}_{i}= D_{i}-m\Gamma_{i} \quad ~ {\rm if} ~ d~ {\rm odd},
\qquad
{\tilde D}_{i}= D_{i}-i m\Gamma_{i}\Gamma_{*}
\quad ~ {\rm if} ~ d~ {\rm even}
\label{conkila}
\end{equation}
  then the cosmological constant changes sign to become negative and the
holonomy group ${\cal H}(\ti D)$ is in $\Spin(d,1)$. The subgroups of  
$\Spin(d,1)$ that give rise to Killing spinors have neen classified in  
\cite{Bryant}.
In this case the generalised holonomy ${\cal H}(\tilde D)$ is the  
holonomy of the Levi-Civita connection on
a timelike cone $\ti C(X)$ over $X$, with metric
$ds^2_{\tilde C(X)}=-dt^2 + t^2 ds^2_X$.
This Killing spinor equation arises in the supergravity theories of  
\cite{Hull:1998vg}
and hyperbolic space $H^d=\SO(d,1)/SO(d)$ is maximally supersymmetric  
with trivial holonomy,
and the cone $\ti C(H^d)$ is $d+1$ dimensional Minkowski space.

Similarly, one can consider the Killing spinor equations for Lorentzian  
spaces $X$ with the Riemannian holonomy contained in $\Spin(d-1,1)$.
In this case, with
supercovariant derivatives
given by  (\ref{conkil}),
  the generalised  holonomy is contained in $\Spin(d-1,2)$, so that the  
maximally supersymmetric case with trivial holonomy is anti-de Sitter  
space; this again arises in many supergravities.
  The holonomy ${\cal H}(\ti D)$ is the holonomy of the Levi-Civita  
connection on
a timelike cone $\ti C(X)$ over $X$, with metric of signature $(d-1,2)$.
  If $m$ is replaced with $im$ to give  (\ref{conkila}), the generalised  
  holonomy is contained in $\Spin(d,1)$ instead, with maximally  
supersymmetric solution given by de Sitter space, as in the  
supergravity theories of  \cite{Hull:1998vg,{Hull:1998ym}}. The
corresponding cone   
is the spacelike cone $C(X)$ over
  $X$, with Lorentzian signature $(d,1)$.

%%%%%%%
    \subsection{Direct Products with Flux}

We will seek backgrounds admitting Killing spinors satisfying
\begin{equation}
\label{fistr}
   \frac{1}{24}F_{MPQR} \Gamma^{MPQR}\epsilon=0
\end{equation}
Such a constraint was used in \cite{Taylor},\cite{Beckera},\cite{Beckerb}.
Then for such solutions, the Killing spinors satisfy  $\hat D  
\epsilon=0$ where
$\hat D$ is the associated derivative
\begin{equation}
{\hat D}_{M} =(D_{M}+ \frac{1}{24} \Gamma^{PQR}F_{MPQR})
  \label{kiln}
\end{equation}
and we will analyse the holonomy of $\hat D$.
  This holonomy group is in the group generated by $\Gamma_{(2)},  
\Gamma_{(3)}$.
In general these generate the whole of $\SL(32,\R)$,  but   further  
assumptions about the configuration lead to interesting restrictions.

Consider product spaces $M_d\times M_{\ti d}$ with a $d/\ti d$ split,  
so that the
coordinates can be split into $x^\mu, y^i$ with $\mu,\nu=1,...,d $ and  
$i,j=1,...,\ti d = 11-d$, and a product
metric of the form
\begin{equation}
g^{(11)}_{MN}=\pmatrix{ g_{\mu\nu}(x)&0\cr0&g_{ij}(y)}
\end{equation}
with one of the metrics    $ g_{\mu\nu} (x) ,g_{ij}(y)$
having Lorentzian signature and the other Euclidean signature.
One of these spaces must be even dimensional; suppose it is $M_{\ti  
d}$. A convenient realisation of the gamma matrices
$\Gamma_M$ in terms of the gamma matrices $\gamma_\mu$ on $M_d$ and the  
ones
$\ti \Gamma_i$ on $M_{\ti d}$ is
\begin{equation}
\Gamma_\mu = \gamma_\mu \otimes  \ti  \Gamma_*, \qquad \Gamma_i = 1  
\otimes \ti  \Gamma_i
\end{equation}
where $\ti  \Gamma_*$ is the chirality operator on $M_{\ti d}$,  $\ti   
\Gamma_*\propto \prod_i  \ti  \Gamma_i $.

The
holonomy of the Levi-Civita connection is in the group generated by  
$\Gamma_{\mu\nu},\Gamma_{ij}$, and so ${\cal H}(D) \subseteq  
\Spin(d-1,1)\times \Spin(\ti d)$
if $M_d$ is Lorentzian, or ${\cal H}(D) \subseteq \Spin(d)\times  
\Spin(\ti d-1,1)$ if $M_{\ti d}$ is Lorentzian.
If the only non-vanishing components of $F$ are $F_{ijkl}$, then
the holonomy of $\ti D$ is in the group generated by
$\Gamma_{\mu\nu},\Gamma_{ij},\Gamma_{ijk}$,
so ${\cal H}(\ti D) \subseteq \Spin(d-1,1)\times G_{spacelike}(\ti d)$
or ${\cal H}(\ti D) \subseteq \Spin(d )\times G_{timelike}(\ti d)$.
Similarly, if
the only non-vanishing components of $F$ are $F_{\mu\nu\rho\sigma}$,  
then
the holonomy of $\ti D$ is in the group generated by
$\Gamma_{\mu\nu},\Gamma_{ij},\Gamma_{\mu\nu\rho}$,
which is $ G_{timelike}(d )\times \Spin(\ti d)$ or $  
G_{spacelike}(d)\times \Spin(\ti d-1,1)$.

Next suppose  both $F_{ijkl}$ and $F_{\mu\nu\rho\sigma}$ are non-zero,  
and all other components are zero.   This requires a 7/4 or 5/6 split.
Then the holonomy is in the group generated by
$\Gamma_{\mu\nu},\Gamma_{ij},\Gamma_{ijk},\Gamma_{\mu\nu\rho}$
and so contains both $ G_{spacelike}(n)$ and $G_{timelike}( 11-n)$  
(where $n$ is the dimension of the spacelike factor), but for 7/4 and  
5/6 splits, these two subgroups do not commute.
For example, the commutator $[\Gamma_{ijk},\Gamma_{\mu\nu\rho}]$  
includes
the term $\Gamma_{ijk \mu\nu\rho}$ and the holonomy in this case is in  
general in $SL(32,\R)$.

Consider further the example of  a 7/4 split, with a Lorentzian 4-space.
On the four-dimensional factor, the maximal structure group is $SO(3,2)$
generated by $\Gamma_{ij},\Gamma_{ijk}$.  The maximal subgroup of  
$SL(32,\R)$ commuting with this is $SU(8)$, generated by
$\Gamma_{\mu\nu} ,\Gamma_{\mu\nu\rho\sigma },  
\Gamma_{\mu_1\mu_2...\mu_6 }$.
However, the 4-gamma term  $\Gamma_{\mu\nu\rho\sigma }$ does not occur  
in the supercovariant derivative, nor does it occur in the commutators  
of terms that do, so that the maximal   structure group containing   
$SO(3,2)$
that is a proper subgroup of $SL(32,\R)$ and can arise as a holonomy of
the supercovariant derivative is $SO(3,2)\times SO(8)$, and
$\SO(3,2)\times \SU(8)\subset \SL(32,\R)$ does not occur.

Similarly, for a 5/6 split  with a Lorentzian 5-space the structure groups
$\SO(4,1)\times USp(8)$ and $\SO(5,\C)  
\times \SO(6)$ are possible (from table 1)
but $\SO(5,\C) \times USp(8)$ is not as it is not contained in  
$\SL(32,\R)$.
Although $\SL(32,\R)$ has subgroups $SO(4,2)\times USp(8)$ and  
$\SO(5,\C) \times \SO(6,\C)$, neither of these arise as structure  
groups as the extra generators are  products of four gamma matrices, which
do not occur in the supercovariant derivative.

  \subsection{Warped Products with Flux}

Consider now a warped product with a 3/8 split, with metric of the form  
(\ref{ansatz}) with $d=3$, and 4-form field strength with $F_{ijkl},
F_{\mu\nu\rho  
i} $ the only non-vanishing components of the 4-form field strength, as  
in \cite{Beckera,Beckerb}.
  Then the Killing spinor $\eta$ can be decomposed  into a 2-component  
$\Spin(2,1)$ spinor $\epsilon$ and a 16-component
  $\Spin(8)$ spinor $ \xi$
  \begin{equation}
\label{ }
\eta=\epsilon \otimes \xi
\end{equation}
and $\xi$ can
be decomposed into 8-component chiral spinors $\xi =\xi_++\xi_-$, with  
$\ti \Gamma_* \xi_\pm =\pm \xi_\pm $.
As in \cite{Beckera,Beckerb}, we consider configurations with
\begin{equation}
\label{ }
F_{\mu\nu\rho i} =\epsilon_{\mu\nu\rho} \partial_i \Delta^{-3/2},
\end{equation}
and
\begin{equation}
\label{ }
F_{mnpq}\gamma^{mnpq} \xi=0
\end{equation}
Then the condition $\ti D_\mu \eta =0$ gives
\begin{equation}
\label{ }
  D_{\mu} \eta   +{1 \over 4} \partial_n ( \log \Delta)
\left[ \gamma_{\mu} \otimes \ti \Gamma^n (1-\ti \Gamma _*) \right] \eta  
=0
\end{equation}
As  $\ti D_\mu$ commutes with $1 \otimes \ti\Gamma_*$, one can decompose
$\eta =\eta _++\eta _-$, with  $\eta_\pm=\epsilon \otimes \xi_\pm$ and  
consider the
action of $\ti D_\mu$ separately on $\eta _\pm$.
On $\eta _+$, the holonomy is ${\cal H}(\ti D_\mu )^+ \subseteq  
\Spin(2,1)\times \Spin(8)$
while on $\eta _-$, the term involving $\partial \Delta$ leads to the  
structure group generated by
$\gamma_{\mu\nu}\otimes 1, 1\otimes \ti \Gamma_{mn}$ and $   
\gamma_{\mu} \otimes \ti \Gamma_n (1-\ti \Gamma _*)$, which is the  
semi-direct product $[\Spin(2,1)\times \Spin(8)]\ltimes
\R^{24}_{(3,8)}$.
Then
${\cal H}(\ti D_\mu)^-  $ is contained in this group.
The Killing spinor condition $\ti D_\mu \eta =0$ is satisfied if   
$\eta=\epsilon \otimes \xi_+$ with chiral $\xi$ and
$D _\mu \epsilon=0$.
If
$  \partial   \Delta =0$, there is no warping and the space is a direct  
product with $F_{\mu\nu\rho i} =0$ and
   the holonomy for $\eta ^+,\eta^-$ is in $\Spin(2,1)\times \Spin(8)$.

The remaining conditions  $\ti D_i \eta =0$ give
\begin{equation}
\label{ }
 D_m \xi
+\frac{1}{24} \Delta^{3/2} \ti \Gamma^{npq} F_{mnpq} \xi +
\frac{1}{ 4}  {\partial}_m (\log \Delta) \xi
-\frac{3}{8} \partial_n (\log \Delta) {\g_m}^n \xi =0
\end{equation}
For this to have a solution with chiral $\xi$ requires
\begin{equation}
\label{ }
\ti \Gamma^{npq} F_{mnpq} \xi=0
\end{equation}
Then the  holonomy of $\ti D_i $ is in $\CSpin(8) = \Spin(8)\times \R^+$
with a conformal piece $ \R^+$, and a Weyl transformation
$g_{ij} \to \hat g_{ij}=\Delta^{-1/2} g_{ij}$ brings   this
to
\begin{equation}
\label{ }
\hat D _i \hat \xi =0
\end{equation}
where $\hat D _i$ is the Levi-Civita connection for $\hat g_{ij}$  
and
$\hat \xi =\Delta^{ 1/4}\xi $, and  ${\cal H}(\hat D_i )  \subseteq   
\Spin(8)$. This requires that $\hat \xi_+$ is covariantly constant with  
respect to $\hat D _i$, and so
$  \hat g_{ij}$ must be a special holonomy metric.
For one parallel spinor on the eight-manifold ${\cal H}(\hat D_i )   
\subseteq  \Spin (7)^+$ and for two
${\cal H}(\hat D_i )  \subseteq SU(4)$.

Then with this ansatz  the
structure group is
${\cal G}=  \CSpin(8)^+\times ( [\Spin(2,1)\times \Spin(8)^-]\ltimes
\R^{24}_{(3,8)})$, where $ \Spin(8)^\pm$ act  on positive or negative  
chirality spinors, and the holonomy is contained in this.
There will be Killing spinors if
   the 3-space is 3-dimensional Minkowski space,  and the 8-manifold is  
conformally
related to a manifold with special holonomy $H\subseteq \Spin (7)^+$.  
Then
the holonomy group is in
${\cal H}=  H\times \R^+ \times (   \Spin(8)^- \ltimes
\R^{24} )$.

  If $\Delta$ is constant so that there is no warping, the structure  
group
  reduces to ${\cal G}=
  \Spin(2,1)\times \Spin(8)^+\times \Spin(8)^-
  $, which is contained in the group ${\cal G}=
  \Spin(2,1)\times \Spin(16)$
   of table 1.
   The holonomy would then be in the subgroup $1\times H\times  
\Spin(8)^- \subseteq  \Spin(2,1)\times \Spin(8)^+\times \Spin(8)^-$.
   However,  for non-trivial warping, one obtains a
holonomy and structure group not contained in any of the groups in  
table 1.

In general there are no negative chirality Killing spinors.
If  $H=\Spin (7)$, one can take $F_{ijkl}$ to be proportional to the  
$\Spin (7)$-invariant 4-form
and
    there are 2 positive chirality  Killing spinors, and  the background  
preserves 1/16 supersymmetry \cite{Beckerb}. If
     $H=SU(4)$, the space is conformal to a Calabi-Yau space and one take
      $F_{ijkl}$ to be a (2,2) form satisfying
       $J^{ij}F_{ijkl}=0$, where $J_{ij}$ is the Kahler form, and
   there are 4    positive chirality  Killing spinors, so the background  
preserves 1/8 supersymmetry \cite{Beckera}.

\subsection{Static Spaces}

Consider static spacetimes
of the form
\begin{equation}
\label{ }
ds^2 =-\Delta(x)dt^2 + g_{ij}dx^i dx^j
\end{equation}
For the ansatz of \cite{Duff:2003ec}, the structure group for $\hat  
D_i$ is in $\SO(32)$, generated by $\{\Gamma_{(2)}, \Gamma_{(3)}\}$,  
which closes on the set of generators
$\{\Gamma_{(2)}, \Gamma_{(3)},\Gamma_{(6)},\Gamma_{(10)}\}$ where
$\Gamma_{(n)}= \Gamma_{i_1...i_n}$ are products of spatial  
gamma-matrices
(so that $\Gamma_{(n)} \propto \Gamma_{(10-n)}\Gamma_0$).
The subset $\{\Gamma_{(2)}, \Gamma_{(6)} \}$
generates $\SU(16)$.
Consider   general electric and magnetic fluxes
$E_{ijk}=F_{0ijk}$ and $B_{ijkl}=F_{ijkl}$.
With general  $E,B$, the structure group is   $\SL(32,\R)$.
For the purely magnetic case, $E=0$,
the structure group for $\ti D_i$  is generated by
$\{\Gamma_{(2)}, \Gamma_{(3)},\Gamma_{(5)}\}$, giving for generic cases  a
holonomy the full  
$\SL(32,\R)$.
However, for those configurations in which the Killing spinor in  
addition
satisfies
\begin{equation}
\label{zxca}
F_{ijkl}\Gamma^{ijkl} \epsilon =0
\end{equation}
the $\Gamma_{(5)}$ term is absent for the corresponding associated
derivative $\hat D$ and the holonomy group ${\cal H}(\hat D)$  is in  
$\SO(32)$.
For $n$ supersymmetries, the holonomy must be in the subgroup  
$\SO(32-n)$.

For electric configurations with $B=0$, the structure group is  
generated by
$\{\Gamma_{(2)}, \Gamma_0\Gamma_{(2)},\Gamma_0\Gamma_{(4)}\}$, which  
closes on the generators
$\{\Gamma_{(2)}, \Gamma_{(4)},\Gamma_{(6)},\Gamma_{(8)} \}$ of  
$\SL(16,\C)$.
For $2n $ supersymmetries, the holonomy must be in $\SL(16-n,\C)$.
If in addition
the Killing spinors
satisfy
\begin{equation}
\label{yyug}
F_{0ijk}\Gamma^{ijk} \epsilon =0
\end{equation}
the $\Gamma_0\Gamma_{(4)}\sim \Gamma_{(6)}$ generator is absent from the
associated connection $\hat D$, so that the holonomy  ${\cal H}(\hat D)$
must be in
the group $\Spin(10,\C)$ generated by $\{\Gamma_{(2)},  
\Gamma_0\Gamma_{(2)}\}$.

In addition one needs to consider the condition
$\ti D_0 \epsilon=0$.
For non-trivial warpings with $\partial _i \Delta\ne 0$, the holonomy  
${\cal H}(\ti D_M)$ will in general be strictly larger than ${\cal  
H}(\ti D_i)$, but in the case of trivial warping in which $\Delta$ is  
constant, the structure groups corresponding to $\ti D_M$ and $\ti D_i$
are the same. 

If the Killing spinor is  
time-independent, $  D_0 \epsilon=0$, then $\ti D_0 \epsilon=0$ becomes  
the  algebraic
equation.
 \begin{equation}
\label{ }
F_{ijkl} \Gamma^{ijkl}\epsilon= -8 F_{0ijk} \Gamma^0\Gamma^{ijk}\epsilon
\end{equation}
If $E=0$, this implies (\ref{zxca}) while if 
$B=0$ it implies (\ref{yyug}). Thus the  conditions
(\ref{zxca}),(\ref{yyug}) naturally arise from requiring $  D_0
\epsilon=0$. If both $E,B$ are non-zero, the holonomy is generic in
general, but if the 10-space has a product structure, then the holonomy
is further restricted and the analysis is similar to that  in section 4.2.

\section{Other $D=11$ Supergravities}
\label{other}

The classical $D=11$ supergravity field equations are invariant under  
the scaling transformations
\begin{equation}
g_{MN} \to \lambda ^2 g_{MN}, \qquad
  \Psi_M   \to \lambda ^{3/2}
   \Psi_M  ,\qquad
A_{MNP} \to \lambda ^3
A_{MNP}
\end{equation}
In \cite{Howe:1997rf}, Howe found a generalisation of the usual $D=11$  
supergravity theory in which this symmetry is made local by coupling to  
a conformal connection $k_M$, which is a gauge field transforming under  
the scaling transformations as $\delta k=d \lambda$.
This requires that the conformal connection be flat, $dk=0$, so that  
locally it is pure gauge and introduces no new degrees of freedom.
However, this generalisation allows new solutions in which $k$ has   
non-trivial holonomy.
A circle compactification with conformal   holonomy around the circle
  (i.e. with a Wilson line for $k$) gives a new massive $D=10 $  
supergravity \cite{Howe:1997qt}
  which has de Sitter solutions \cite{Chamblin:2001dx}.
  In particular, the condition for supersymmetry of  a bosonic  
background becomes
\begin{equation}
({\tilde D}_{M}+k_M)\epsilon=0,
\end{equation}
  which can be analysed in terms of the holonomy of the connection  
${\tilde D}_{M}+k_M$.
  Whereas ${\tilde D}$ takes values in $\SL(32,\R)$, adding the  
conformal connection means that   ${\tilde D}+k$  takes values in  
$\GL(32,\R)$, and the holonomy is a subgroup of
    $\GL(32,\R)$. Thus including the conformal connection allows more  
general configurations.
    For a configuration preserving $n$ supersymmetries, the holonomy of  
${\tilde D}+k$ must be in
$\GL(32-n,\R)\ltimes \R^{n(32-n)}$. All values of $n$ are   possible in
principle, as a non-trivial  
holonomy
of $GL(1,\R)\ltimes \R^{31}$   allows $n=31$ supersymmetries, as would a
holonomy in the subgroup $GL(1,\R)$.

In\cite{Chamblin:2001dx}, it was suggested that this modified $D=11$  
supergravity might
arise as a limit of a modified M-theory, referred   to as MM-theory.
  However, the classical scaling symmetry of the usual $D=11$  
supergravity is not a symmetry of the quantum theory. For example, the  
supergravity field equations   receive  higher derivative corrections  
in the quantum theory, and these break the scaling symmetry as  each    
higher derivative term will scale according to the number of  
derivatives.
  Then it would be inconsistent to gauge the scaling  symmetry in the  
quantum theory,
  so that it would seem that M-theory could not be a part of an   
MM-theory.
  (However, such a structure could be of interest if M-theory had a  
scale invariant phase.)

  In \cite{Hull:1998ym}, it was shown that in addition to the classical  
supergravity in 10+1 dimensions, there are supergravities in 9+2 or 6+5  
dimensions, and these signatures with 1,2 or 5 times  are the only  
possibilities that can arise in eleven dimensions (together with  the  
mirror theories in 1+10, 2+9 or 5+6 dimensions). Chains of dualities  
involving solutions with periodic time
\cite{Hull:1998ym}   lead to phases of M-theory in 9+2 or 6+5  
dimensions, and the supergravity theories arise as limits of these.
    The arguments leading to these exotic phases are formal and assume  
that the quantum theory is consistent in configurations with periodic  
time.
   The supergravities are similar in structure to the usual one, and
   supersymmetric   solutions   have been found in   \cite{Hull:1998fh},  
\cite{Hull:1999mt}.
The conditions for Killing spinors
   can be analysed in terms of the holonomy of  a supercovariant  
derivative of the same form as
   $\tilde D$. Similar groups to those in table 1 arise as possible  
holonomies (they are different real forms of the same complex groups),  
but the general holonomy is again  $\SL(32,\R)$.
   The subgroup generated by $\Gamma_{(1)},\Gamma_{(2)}$
  is $\Spin (10,2)$ or $\Spin(6,6)$ in the two cases.
  It is interesting that a formulation of M-theory with local  
$\SL(32,\R)$ symmetry could be
  a natural framework to incorporate the conjectured phases with  
signatures 9+2 and 6+5, together with the theory in 10+1.

\section{M-Theory}
\label{nosusy}

In $d=11$ supergravity, the fermion fields are sections
of the spin bundle, with transition functions in $\Spin(10,1)$, and the  
number of supersymmetries preserved by a background depends on the number
of solutions  
to the Killing spinor condition.
In M-theory, there are   vacua that do not correspond to supergravity  
solutions. More surprisingly, there are supergravity solutions that are  
also solutions of M-theory, and which are known to be supersymmetric   
vacua of M-theory but for which the supergravity solution has no  
Killing spinors, or fewer Killing spinors than the number of expected  
supersymmetries \cite{DLP1,DLP2,Pope:1999xg}.
This is the phenomenon of \lq supersymmetry without supersymmetry', and  
brane wrapping modes
or non-perturbative string states
play a crucial role in realising the supersymmetry in such cases.

An example illustrating this is obtained as follows \cite{DLP2}.
Consider the $AdS_5\times S^5$ solution of the type IIB string theory,  
which is maximally supersymmetric with 32 Killing spinors and has an RR  
5-form flux.
The 5-sphere  admits a Hopf fibration as an $S^1$ bundle over $\CP^2$.  
The isometry along the $S^1$ can be used to perform a T-duality taking  
this to a solution of the IIA string theory
in which the bundle is untwisted to give the product space
   $\CP^2\times S^1$ with the Fubini-Study metric on $\CP^2$, and with a  
NS-NS 2-form field now turned on. The number of supersymmetries is  
expected to be preserved by the T-duality, but surprisingly the IIA  
solution not only does not have any Killing spinors, it does not have  
any spinors
   at all, as   $\CP^2$ does not admit a spin structure. The IIA theory on
$AdS_5\times\CP^2\times S^1$ 
    is supposed to give a dual description of the IIB string theory  in
the $AdS_5\times S^5$ vacuum, so the question arises as to what happened
to the IIB fermions.
   The resolution is that all the spinors on $S^5$ have non-trivial  
dependence on the $S^1$ direction and so can be thought of as carrying  
momentum in that direction, so that in the T-dual picture the spinors  
of the original theory all now arise in the winding sector, and there  
are no fermions at all in the zero-winding sector.

   This can be understood through the dimensional reduction to $d=9$  
\cite{DLP2}.
    Reducing the $AdS_5\times S^5$ solution of the type IIB supergravity  
along the Hopf fibre direction gives a solution of $d=9$ supergravity  
on $AdS_5\times \CP^2$
    with a 2-form flux $F$ proportional to the Kahler 2-form on $\CP^2$
and a 4-form flux proportional to the volume form on $\CP^2$.
Here $F=dA$ and $A$ is the Kaluza-Klein vector field coming from
the reduction of the metric.
The dimensional reduction of the $d=10$ fermion fields gives
$d=9$ fermions which are all charged with respect to the $U(1)$ and so  
couple to $A$.
However, the fermions are not $d=9$ spinors, i.e. they are not sections  
of a $\Spin(8,1)$ bundle over $AdS_5\times \CP^2$, because $\CP^2$ does  
not admit a spin structure.
However, $\CP^2$ does admit a spin${}^c$ structure or generalised spin  
structure which allows charged spinors, arising as sections of a  
$\Spin(8,1)\times U(1)$ bundle. Whereas there can be no spinors  
coupling just to the spin conection $\omega$, there can be spinors  
coupling to
the combined connection $\omega+A$, provided that the $U(1)$ charge is  
half-integral.
The fact that the charge is half-integral instead of the integral  
charge usually required
  gives an extra minus sign that cancels the sign inconsistencies that
arise  
in attempting to define spinors on the manifold.
  The supersymmetry parameters $\epsilon$ are also charged spinors that
are
sections of the $\Spin(8,1)\times U(1)$ bundle
  and the Killing spinor condition involves the   combined connection  
$\omega+A$.
  The IIB Killing spinors in $d=10$ give rise to charged Killing spinors  
in $d=9$.

  All spinor fields on  $AdS_5\times S^5$  give rise to charged spinor  
fields
  on  $AdS_5\times \CP^2$ coupling to
   $\omega+A$, and there can be no uncharged spinors.
   Lifting the $d=9$ solution up to
  the $AdS_5\times \CP^2\times S^1$ solution of the IIA theory, the  
vector field $A$ lifts to the NS-NS 2-form $B_2$ and so all the spinor  
fields of the IIB theory have become winding modes of the fundamental  
IIA string on the $S^1$ coupling to $B_2$, and again there are no such  
fields that do not wind. There are no gravitini or spin-half fermions  
arising as fields on
    $AdS_5\times \CP^2\times S^1$ (as there can be no spinor fields) but  
the fermionic states arise in the winding sector. Just as the charge in  
$d=9$ was half-integral, the charges governing the coupling to $B_2$  
are half-integral, so that one might say that  the fermion states have  
half-integral \lq winding number' or string charge, or that they are  
fractional strings.
    In particular, the supersymmetry parameters $\epsilon$ are not  
spinor  fields but have
    half-integral  winding number. As the string charges are all
half-integral, there is no zero-winding sector.

    Next, the $AdS_5\times \CP^2\times S^1$ solution of the IIA theory  
can be lifted to
$AdS_5\times \CP^2\times T^2$ solution of M theory \cite{DLP2}.
 The string winding
     modes have become modes coupling to $A_3$ and so might be thought  
of as membrane wrapping modes, \lq wrapping' the
     $T^2$, although the \lq wrapping number'  or membrane charge is  
half-integral, so that they are fractional membranes.

    The $AdS_5\times S^5$ solution has isometry group $SO(4,2)\times  
SO(6)$, and the $SO(6)$
    gives rise to an
    $SO(6)$ Yang-Mills symmetry on reducing to five dimensions.
    The $d=9$ solution $AdS_5\times \CP^2$
    has an internal space with isometry of only
    $SU(3)\times U(1) $, so that this will be the Kaluza-Klein gauge  
symmetry on reducing from nine to  five dimensions.
    The remaining gauge fields of $SO(6)$ arise from charged fields in  
$d=9$, which lift to winding modes in the IIA theory or membrane  
wrapping modes in 11-dimensions \cite{DLP2}.
    On the other hand, the 11-dimensional solution $AdS_5\times  
\CP^2\times T^2$ has internal space   isometry
    $SU(3)\times U(1)^3 $, so that this is the Kaluza-Klein gauge  
symmetry on reduction to five dimensions.
    The extra $U(1)^2$ gauge fields arise from massive modes on  
$AdS_5\times S^5$ in the IIB picture \cite{DLP2}.

 Dualities can give rise to \lq  wrapping modes'  coupling to  other  
form fields. As another example, consider the $AdS_4\times S^7$ solution
of  
M-theory.
  The 7-sphere is a Hopf fibration of $S^1$ over $\CP^3$, so reducing on  
the $S^1$ fibre  will give
  a IIA solution $AdS_4\times \CP^3$ \cite{DLP1}. The fermions
  give d=10 fields that couple to the Kaluza-Klein vector field, which  
is the RR gauge field of the IIA theory, so that they carry RR charge.
  Now a series of $p$ T-dualities
  give rise to fermion fields coupling to the RR $p+1$ form gauge field  
$C_{p+1}$ and so could be
  said to carry D-brane charge or to  be D-brane wrapping modes.

In   standard supergravity theories, the fermionic fields are spinors. A
conventional viewpoint would be to say that backgrounds without spin
structure are forbidden as configurations of the theory. Another view
would be to allow non-spin   solutions, finding that that there are no
fermions in  the spectrum of fluctuations about the background.
Alternatively, the supergravity  theory can be modified 
for backgrounds with a spin${}^c$ structure
to allow fermions that are charged spinors. The key point here is that we
have learnt that in M-theory the
supergravity limit is indeed modified in precisely this way and that the
fermion fields are charged spinors in general, so that backgrounds that
are not spin but which have
a spin${}^c$ structure are indeed allowed.
 
  Let us return to the  $AdS_5\times \CP^2$ solution of $d=9$  
supergravity.
  There M-theory requires that the supergravity is modified so that the
fermions are not spinors but are charged fields arising as  
sections of a bundle with $\Spin(8,1)\times U(1)$  transition functions.
  However, $\Spin(8,1)\times U(1)$ is precisely the structure group  
${\cal G}=\Spin(8,1)\times G_{spacelike}(2)$  appropriate for a 9/2  
split. The local $\Spin(8,1)\times U(1)$ symmetry is then crucial for the  
definition of the theory, and the physical  fermionic fields  arise as  
sections of bundles with this structure group.
  This is an important  piece of evidence that enlarged structure groups  
should play an essential  role in M-theory.
   Similar arguments lead to other structure groups such as those in table
1 arising
   with \lq charged spinors' that are sections of $ {\cal G}$-bundles,  
suggesting that the general picture should involve bundles over  
spacetime with $\SL(32,\R)$ transition functions.
All the examples that arise have transition functions in $ {\cal
G}\subseteq \SL(32,\R)$
for various $ {\cal G}$ so all are particular $\SL(32,\R)$-bundles.
Just as in coupling   spinors to gravity one needs to use a formulation
with local Lorentz symmetry, to encompass  charged spinors 
that are sections of a ${\cal G}$-bundle
requires a formulation with local ${\cal G}$ symmetry, with different
${\cal G}$ for different backgrounds. 
A formulation with local $\SL(32,\R)$ symmetry would allow all possible
$ {\cal G}\subseteq \SL(32,\R)$ bundles without 
having to specify a background, and seems to be the minimal requirement
for a background-independent formulation. That would mean that  the  
32-component indices $\alpha,\beta$ should be regarded not as
$\Spin(10,1)$
  spinor indices but as $\SL(32,\R)$ indices in the fundamental  
representation.

Some care is needed in this discussion, as there are a number of related
but distinct  symmetries that play a role here.  The $d=9$ theory arising
from the Hopf reduction 
can be viewed as a gauged supergravity, and many related constructions
lead to gauged supergravities. In such theories, the fermions    are
typically spinors transforming under the gauge group coupling both to the
spin connection and the gauge connection and   M-theory allows fermions
that are not spinors but are
sections of a spin${}^c$-bundle.
A more detailed discussion of the relevant symmetries from the point of
view of gauged supergravity is given in an appendix.

M-theory in a particular background then requires that the   local Lorentz
symmetry
  be extended to    at least a local
  ${\cal G}$ symmetry, where the group ${\cal G}$ depends on the
background.
   Extending  to  a local $\SL(32,\R)$ symmetry 
   removes dependence on the choice of background and includes all
possible 
   ${\cal G} \subseteq \SL(32,\R)$ symmetries. A reformulation of $d=11$
supergravity with local
   $ \SL(32,\R)$ symmetry would be a useful first step towards 
   such a formulation of M-theory.

    The $d=11$ superalgebra can be written as
     $\{ Q_\alpha , Q_\beta \}= \Pi_{\alpha  \beta }$ where the
     symmetric bi-spinor $\Pi _{\alpha  \beta }=
     P_M \Gamma ^M_{\alpha  \beta } + Z_{MN}\Gamma ^{MN}_{\alpha  \beta  
}+... $ can be decomposed in terms of the 11-momentum $P_M$, a membrane  
charge and other brane charges \cite{Hull:1997kt}.
     If the indices  $\alpha$ are thought of as $\SL(32,\R)$ indices,  
then there is no invariant way of
     making this decomposition, which
      requires choosing a $\Spin(10,1)$ subgroup of $\SL(32,\R)$.
      In other words, an   $\SL(32,\R)$ symmetry or enlarged symmetry $  
{\cal G}\subset \SL (32,\R)$ would mix the momenta with brane and other  
charges.

   However, this cannot be the whole   story. In addition to  
degrees of freedom that are spacetime  fields  or  sections of  
${\cal G}$-bundles over spacetime, there are string winding modes and
brane wrapping  
modes (sometimes with fractional charges or winding numbers)
   that play a crucial role and which should be taken into account when  
considering questions of supersymmetry and  symmetry of any given  
vacuum.
   It is not known what the right formulation is for properly  
considering such modes in M-theory.
   One approach for string winding modes might be to replace space with  
a loop-space (as in string field theory), or with the space of maps from a
$p$-dimensional space to  
spacetime for $p$-brane wrapping modes.
   Another possibility  that has been considered is to extend spacetime  
with extra   coordinates conjugate to brane charges as well as the  
usual ones that are conjugate to momenta.
          One could then have 528 coordinates $  \Xi^{\alpha  \beta }$  
conjugate to
  the charges     $  \Pi_{\alpha  \beta }$, with the usual spacetime  
arising as an 11-dimensional subspace or \lq brane'   in this large  
space. A duality transformation would then take one to a different  
11-dimensional subspace, so that what was previously a brane charge  
becomes a momentum and vice versa. The dualities of  
\cite{Hull:1998ym} could be incorporated in such a picture.
  In addition, one could introduce a fermionic coordinate  
$\Theta^\alpha$, with $\SL(32,\R)$ acting naturally on the superspace   
with coordinates $  \{ \Xi^{\alpha  \beta }, \Theta^\alpha \}$.
Many serious problems arise in attempting such   formulations, but they  
perhaps deserves further exploration.

\section*{Appendix: Charged Spinors and Gauged Supergravity}

It will be useful to compare the $d=9$ theory appearing in the Hopf
reduction of $AdS_5\times S^5$  with the usual $d=9$ supergravity.
The $d=9$ supergravity from standard toroidal reduction   from 11
dimensions has three abelian vector gauge fields and so a local $U(1)^3$
symmetry,
and three scalars taking values in $\R^+\times \SL(2,\R)/U(1)$. The theory
has a nonlinearly realised
 $\R^+\times \SL(2,\R)$ global symmetry. This becomes linearly realised on
introducing an extra scalar and an extra local $U(1)$ symmetry, which will
be denoted $U(1)_F$. The theory then has $\R^+\times \SL(2,\R)$ global and
$U(1)^4$ local symmetry. The fermions are charged only with respect to the
extra $U(1)_F$, and couple to a $U(1)_F$ composite connection ${\cal B}_M$
constructed from the scalar fields; this is not an independent gauge
field.
 On fixing the extra $U(1)_F$ symmetry by setting the extra scalar to
zero, the
 $\R^+\times \SL(2,\R)$  acts on the fermions through a compensating
$U(1)_F$ transformation, needed to preserve the gauge condition.

The $d=9$ theory that arises by Hopf reduction is a gauged version of this
supergravity in which
a $U(1)_G\subset \R^+\times \SL(2,\R)$ global symmetry is promoted to a
local symmetry,
with the corresponding gauge field the Kaluza-Klein vector field $A_M$.
(It is presumably  one of the gauged supergravities discussed in
\cite{Hull:2002wg, Bergshoeff:2002mb,Bergshoeff:2002nv}.)
In the formulation with local  $U(1)_F$, the fermions do not transform
under
$U(1)_G$, but now the $U(1)_F$ connection ${\cal B}_M$ depends on the
gauge connection $A_M$ as well as on the scalars, so that ${\cal
B}_M=A_M+(scalar-dependent ~terms)$. On going to physical gauge by setting
the extra scalar to zero, any 
local $U(1)_G$ transformation is accompanied by a compensating 
 $U(1)_F$ transformation which does act on the fermions, so that the
 two $U(1)$'s effectively become identified (the gauge symmetry being now
a diagonal subgroup of $U(1)_G\times U(1)_F$).
 Independently of whether the gauge is fixed or not, the fermions strictly
speaking couple
to $\omega_M+{\cal B}_M$, not $\omega_M+A_M$, and these two differ by
scalar dependent terms. The fermions are sections of the $\Spin(8,1)\times
U(1)_F$ bundle, and the Hopf fibration implies that this must be
non-trivial.

A similar situation arises more generally.
In dimensionally reduced supergravity theories in $d=11-n$ dimensions,
there is a global $\E_{n(+n)}$ symmetry with physical scalars taking
values in $\E_{n(+n)}/G(n)$.
Introducing extra scalars taking values in $G(n)$, the complete set of
scalars now take values in
$\E_{n(+n)}$ 
and there is a local
$G(n)$ symmetry. The fermions
are charged under $G(n)$ but are $\E_{n(+n)}$  singlets.
When the $G(n)$ symmetry is fixed by eliminating the extra scalars, the 
$\E_{n(+n)}$  symmetry acts on the fermions through a compensating  $G(n)$
transformation.
In a gauged version of the theory, a subgroup $K\subseteq \E_{n(+n)}$ is
promoted to a local symmetry, with vector fields from the supergravity
theory becoming the $K$ gauge fields.\footnote{In the $d=9$ example,
$E_{2(+2)}=\R^+\times \SL(2,\R)$, $G(2)=U(1)_F$ and $K=U(1)_G$, while for
gauged supergravities in $d=4$ dimensions with $n=7$,
$G_{spacelike}(7)=SU(8)$ and the gauge group could be $K=SO(8)$ as in
\cite{deWit:1981eq} or $K=SO(p,8-p)$ as in \cite{Hull:vg , Hull:qz}. For a
timelike reduction to $d=4$, $G_{timelike}(7)=SU^*(8)$, and the \lq
natural gauging' analogous to the $SO(8)$ gauging of  \cite{deWit:1981eq}
is   one with gauge group $SO^*(8)=SO(6,2)$ arising from a consistent
truncation of the dimensional  reduction of the $AdS_7\times S^4$ solution
on $AdS_7$. Other gaugings are also possible.}
The fermions couple to a composite $G(n)$ connection
${\cal B}_M$ which depends on the
$K$ gauge connection $A_M$ as well as on the scalars, and so are sections
(for a spacelike reduction) of a
${\cal G}=\Spin(d-1,1)\times G_{spacelike}(n)$ 
 bundle. However,   typically the transition functions can be taken to be
in $K_c\subseteq G(n)$ where $K_c $ is the maximal compact subgroup of
$K$, so that the bundle has a $K_c $ structure.\footnote{For a timelike
reduction, $K_c$ is again a maximal subgroup of $K$, but is now typically
non-compact; it is the maximal subgroup of $K$ that is also a subgroup of
$G_{timelike}(n)$.}
 The fermions are charged and are
   sections of a ${\cal G}=\Spin(d-1,1)\times G_{spacelike}(n)$ bundle,
and a local ${\cal G}$ symmetry (or at least a local $ \Spin(d-1,1)\times
K_c$
symmetry)    is needed to formulate the theory.
 
 Then in a gauged supergravity with a gauge group $K$, fermions   are
typically charged under the gauge group, which acts on the fermions
through $G(n)$ transformations. In general, a spin${}^c$ structure is to
be expected, with the fermions arising not as spinors but as sections of a
bundle with transition functions in ${\cal G}$, which is
$\Spin(d-1,1)\times G_{spacelike}(n)$
 for spacelike reductions. Non-trivial spin${}^c$  structures will often
arise from reduction  of
 situations with supersymmetry without supersymmetry, as in the example
above.
 Strictly speaking the transition functions are in
 $\Spin(d-1,1)\times K$, with 
 $K$ acting through compensating
 $G(n)$ transformations and 
 $K_c\subseteq G(n)$ acting linearly.
 However, it is useful to think of these as $\Spin(d-1,1)\times
G_{spacelike}(n)$ bundles, and this allows the gaugings with different
gauge groups $K$ to be treated on the same footing, although in each case
the bundle can be viewed as
 a $\Spin(d-1,1)\times K_c$ bundle.
  In this way, fermions can arise as sections of ${\cal G}$-bundles 
   for   various structure groups in table 1, and the theory is naturally
formulated with local 
    ${\cal G}$ symmetry.
   If M-theory is to have a background-independent  formulation that is
independent of 
   $d$ and the choice of $K$, then the local symmetry must be one that
includes the various groups 
    ${\cal G}$  that can arise in this way.
    This would require a formulation with local $\SL(32,\R)$ symmetry,
with
    fermions arising  as sections of $\SL(32,\R)$
 bundles. In particular cases, the bundle often reduces to one with
transition functions in a subgroup ${\cal G}$, such as the group
 $\Spin(d-1,1)\times K_c\subset SL(32,\R)$ for the gauged supergravities.

%%%%%%%%%%%%%%%%%%%%%%%%%%%%%%%%%%%%%%%%%%%%%%%%%%

\section*{Acknowledgments}

I would like to thank Dan Waldram and Hermann Nicolai for useful  
discussions, and Jose Figueroa-O'Farrill for pointing out an error in an
earlier version.

%%%%%%%%%%%%%%%%%%%%%%%%%%%%%%%%%%%%%%%%%%%%%%%%%%

%%%%%%%%%%%%%%%%%%%%%%%%%%%%%%%%%%%%%%%%%%%%%%%%%%%%%%%%%%%%%%%%%%
\end{document}